\documentclass[english,aps,manuscript]{paper}
\usepackage[T1]{fontenc}
\usepackage[latin9]{inputenc}
\usepackage{amsmath}
\usepackage{amssymb}
\usepackage{esint}

\makeatletter
\usepackage{babel}

\makeatother

\usepackage{babel}
\begin{document}

\title{Exact, stable, two-derivative interacting massless multi-graviton
theories}

\author{Idan Talshir}

\maketitle
idan.talshir@mail.huji.ac.il

School of Physics, Tel Aviv University, Ramat Aviv 69978, Israel

Department of Physics, Ariel University, Ariel 40700, Israel

Department of physics, Hebrew University, Jerusalem, Israel
\begin{abstract}
We present a general model of interacting metric fields with the sum
of massless non-interacting spin 2 fields as the linear limit. In
the non-interacting limit the model is reduced to a sum of general
relativity actions, with the usual Einstein-Hilbert kinetic term for
each metric field.\\
 Until now the accepted view was that such theories are inconsistent
since it has been proven that ghost terms and/or discontinuity in
the number of degrees of freedom at the zero interaction limit, are
unavoidable. We do not refute these results, but instead prove by
construction that they do not necessarily lead to inconsistencies.
In particular, our suggested theories are:\\
 1. Energetically stable, i.e. their Hamiltonian is bounded from below
and the multi-Minkowski metric configuration is the unique ground
state. 2. Continuous in the zero interaction limit so that general
relativity solutions are restored smoothly. 3. Include no higher than
two derivative Lagrangian terms.\\
 In addition, the dominant energy condition is maintained with respect
to all metric fields for all field configurations.
\end{abstract}

\section{Introduction}

Multi graviton theories are spin two massless fields that without
interaction are described by a sum of Pauli-Fierz actions, which is
a linearized form of a sum of Einstein-Hilbert actions, the free action
for general relativity. The interaction involves, coupling of different
metric fields and their derivatives, beside the usual metric coupling
to standard matter. The interaction is restricted to maximum two derivative
terms, i.e. addends with two first derivatives or one second derivative
in order to keep the classical form of second order differential equations.

These theories were considered not consistent for a long time due
to the impossibility to construct such interacting systems which avoid
negative-propagating excitations and at the same time preserving the
full gauge symmetry of the free limit \cite{inconsistency}. 

The appearance of ghost at some perturbation level may indicate energy
instability. However, the absence of ghost degrees of freedom at any
level of perturbation is not a necessary condition for stability. 

In this paper we construct an energetically stable action that maintains
the conditions of a generalized positive energy theorem \cite{key-9},
an expansion of the general relativity positive energy theorem\cite{key-11}
for several interacting metrics. 

Breaking the symmetry of the free theory may lead to what appears
to be a discontinuity of the solutions in the zero interaction limit,
especially if the limit is  taken in a finite perturbation order of
the full theory. Our model is presented in an exact non-perturbative
form, and the continuity at the zero interaction limit is guaranteed
automatically. This can in principle be shown for all field equations
by explicitly observing that no singularities occur, but it is sufficient
to take a global point of view. In general, the Cauchy problem is
well posed for multi-metric theories\cite{key-4}. Then, propagation
in time of the initial conditions is continuous in the interaction
parameter. When this parameter goes to zero we get identity of some
of the variables that are functionally independent for any finite
value of the interaction parameter, so there are no conceptual problems
concerning with elimination of degrees of freedom.

It should be noted that there are known models with interacting multi-graviton
fields of which only one is massless, and therefore do not fit to
our framework. These models include special interactions which yield
a constraint that removes the ghost degree of freedom.\cite{key-2}

Recently \cite{key-8} we presented a model of energetically stable
bimetric theories, for which the dominant energy condition is valid.
In this paper we construct the action with two derivatives at the
most and explain the continuity at the free limit, therefore provide
a consistent interacting massless multi-metric theory. we also generalize
these results to any number of cross-interacting metric fields and
state the connection to immunity from causality violations.

\section{The model}

The general action is:

\begin{equation}
S=\int{\mathcal{L}{d^{4}}x=\int{\left({\sum\limits _{a=1}^{N}{\alpha^{a}\mathcal{L}_{G}^{a}+{\mathcal{L}_{I}}}}\right)}}{d^{4}}x\label{eq:1}
\end{equation}

where $\alpha^{a}$ are non-negative constants, $\mathcal{L}_{G}^{a}=\frac{1}{16\pi G}{R^{a}}\sqrt{{g^{a}}}$
and ${\mathcal{L}_{I}}={\mathcal{L}_{I}}[g_{\mu\nu}^{1},...,g_{\mu\nu}^{N},{\psi},{\nabla^{1}}{\psi},...{\nabla^{N}}{\psi}]$.
The scalar curvature $R^{a}$ is derived from the metric $g_{\mu\nu}^{a}$
and ${g^{a}}\equiv-\det[g_{\mu\nu}^{a}]$. The fields $\psi$ may
be tensors of any rank, and $\nabla^{a}$ is the covariant derivative
with respect to the metric $g_{\mu\nu}^{a}$. Any standard matter
Lagrangian density is included in the interaction.

The numerical value of the constrained Hamiltonian can be calculated
by any constrained Hamiltonian on the constraint surfaces, without
solving the constraints explicitly. A constrained Hamiltonian density
is the zero-zero component of the Noether current we get from invariance
with respect to coordinate translation:

\begin{equation}
\Theta_{\beta}^{\alpha}=-\mathcal{L}\delta_{\beta}^{\alpha}+\sum\limits _{a}{{g^{a}}_{lm,\alpha}\frac{{\partial\mathcal{L}}}{{\partial{g^{a}}_{lm,\beta}}}}+{{\psi}_{,\alpha}\frac{{\partial L}}{{\partial{\psi}_{,\beta}}}}
\end{equation}

This is the canonical energy momentum pseudo-tensor for the system.
It can be separated to the sum of gravity energy momentum pseudo-tensors
for each metric, and an interaction term for which its space integration
is equal to the space integration of the sum of all energy momentum
tensors for each metric \cite{key-9}

\begin{equation}
\int{\left({-{\mathcal{L}_{I}}\delta_{\alpha}^{0}+\left(\sum_{a}{g^{a}}_{lm,\alpha}\frac{{\partial{\mathcal{L}_{I}}}}{{{\partial^{a}}{g_{lm,0}}}}\right)+{\psi_{,\alpha}}\frac{{\partial{\mathcal{L}_{I}}}}{{\partial{\psi_{,0}}}}}\right)}{d^{3}}x=\int\sum_{a}{\sqrt{-{g^{a}}}{T^{a}}_{\alpha}^{0}{d^{3}}x}
\end{equation}

where

\begin{equation}
{T^{a}}_{\alpha}^{\beta}\equiv{g^{a}}_{\alpha\nu}\frac{2}{{\sqrt{-{g^{a}}}}}\frac{{\delta{\mathcal{L}_{I}}}}{{\delta{g^{a}}_{\nu\beta}}}
\end{equation}

Using field equations and the Belinfante procedure for each metric
\cite{key-10-1} $a$, we get that up to space integration

\begin{equation}
{\Theta^{\mu\nu}}=\frac{1}{{16\pi G}}\frac{\partial}{{\partial{x^{\alpha}}}}\frac{\partial}{{\partial{x^{\beta}}}}\sum\limits _{a}\alpha^{a}{\sqrt{-{g^{a}}}({\eta^{\mu\nu}}{g^{a}}^{\alpha\beta}-{\eta^{\alpha\nu}}{g^{a\mu\beta}}+{\eta^{\alpha\beta}}{g^{a\mu\nu}}-{\eta^{\mu\beta}}{g^{a\alpha\nu}})}
\end{equation}

That is, the total energy momentuom tensor for the interacting multi-graviton
system can be presented as a sum of divergence-free pseudo-tensors,
each of which is a functional of only one metric. If the metric fields
obeys the usual asymptotic conditions

\begin{equation}
g_{\mu\nu}^{a}\simeq{\eta_{\mu\nu}}+{\text{O}}({r^{-1}})\label{eq:6}
\end{equation}

then the canonical energy is equal to a sum of ADM energy \cite{key-10}
expressions:

\begin{equation}
{P^{\nu}}=\sum\limits _{a}{P^{a\nu}}
\end{equation}

where

\begin{equation}
{P^{a0}}\equiv\frac{1}{{16\pi G}}\int{\left({\frac{{\partial{g^{a}}_{ij}}}{{\partial{x^{j}}}}-\frac{{\partial{g^{a}}_{jj}}}{{\partial{x^{i}}}}}\right)}d{s^{i}}
\end{equation}

and

\begin{equation}
{P^{aj}}\equiv\frac{1}{{16\pi G}}\int{\left({\frac{{\partial{g^{a}}_{kk}}}{{\partial{x^{0}}}}\frac{{\partial{g^{a}}_{k0}}}{{\partial{x^{k}}}}{\delta_{ij}}+\frac{{\partial{g^{a}}_{j0}}}{{\partial{x^{i}}}}-\frac{{\partial{g^{a}}_{ij}}}{{\partial{x^{0}}}}}\right)d{s^{i}}}
\end{equation}

The dominant energy condition (DEC) is defined by the requirement
that $T_{\mu\nu}^{a}{u^{a\mu}}{v^{a\nu}}\geqslant0$ for all time-like
future-pointing vectors $u^{a\alpha},v^{a\alpha}$ , where timelikeness
is defined with respect to the metric $g_{\mu\nu}^{a}$ .If DEC is
valid for all energy momentum tensors then the positive energy theorem
\cite{key-11,key-12} can be applied for every index $"a"$ and for
the total energy momentum vector

\begin{equation}
{P^{a0}}\geqslant\left|{P^{ai}}\right|\Rightarrow{P^{0}}\geqslant\left|{P^{i}}\right|
\end{equation}

where equality is obtained when all the metric fields take the Minkowski
matrix form and all (interaction) energy momentum tensors are zero.

We can guarantee DEC for every field configuration if the energy momentum
tensors are conformal to their corresponding metrics

\begin{equation}
T_{\mu\nu}^{a}={F^{a}}g_{\mu\nu}^{a}\label{eq:11}
\end{equation}

and the scalars $F^{a}$ are non-negative for all field configurations

\begin{equation}
{F^{a}}[g_{\mu\nu}^{1},...,g_{\mu\nu}^{N},{\psi},{\nabla^{1}}{\psi},...{\nabla^{N}}{\psi}]\geqslant0
\end{equation}

We now focus on the case where the interaction has no explicit dependence
on metric derivatives.

The energy momentum tensors must be integrated to the same interaction
Lagrangian density, so

\begin{equation}
\frac{{{\delta^{2}}{\mathcal{L}_{I}}}}{{\delta g_{\mu\nu}^{a}\delta g_{\mu'\nu'}^{b}}}=\frac{{\partial\sqrt{{g^{a}}}T_{\mu\nu}^{a}}}{{\partial{g^{b\mu'\nu'}}}}=\frac{{\partial\sqrt{{g^{b}}}T_{\mu'\nu'}^{b}}}{{\partial{g^{a\mu\nu}}}}
\end{equation}

Because derivation of a metric deteriminant with respect to the metric
is conformal to the metric, we choose the dependence of the interaction
on the metric to be solely through the determinants.

One option is to choose energy momentum tensors which are symmetric
in the metric fields

\begin{equation}
\begin{gathered}{F^{a\in X}}={\left({\sum\limits _{b\in X}{\sqrt{{g^{b}}}}}\right)^{-4k}}{\mathcal{D}[{\psi},\psi_{,\rho}]}\\
{F^{a\notin X}}=0
\end{gathered}
\label{eq:14}
\end{equation}

For some set $X\subset\{1,...,N\}$. The constant $k$ is a positive
integer and the functional $\mathcal{D}$ is a non-negative scalar
density with the weight of $4k$..

Integration of the energy momentum tensors gives:

\begin{equation}
{\mathcal{L}_{I}^{(X,k)}}=\int{\frac{{\sqrt{{g^{a\in X}}}}}{2}T_{\mu\nu}^{a}d{g^{a\mu\nu}}=\frac{{\mathcal{D}[{\psi},\psi_{,\rho}]}}{1-4k}}{\left({\sum\limits _{b\in X}{\sqrt{{g^{b}}}}}\right)^{-4k+1}}\label{eq:15}
\end{equation}

This interaction can be generalized by a linear combination of the
determinants with non-negative coefficients summation of all these
possible interactions

\begin{equation}
\mathcal{L}_{I}^{k}=\mathcal{D}[\psi,{\psi_{,\rho}}]\int\limits _{0}^{\infty}{f\left[{{\beta^{1}},...,{\beta^{N}}}\right]}{\left({\sum\limits _{a}{{\beta^{a}}\sqrt{{g^{b}}}}}\right)^{-4k+1}}\prod\limits _{a}{d{\beta^{a}}}\label{eq:15a}
\end{equation}

The non-negative scalar density $D$ may be an even power of the determinant
of the matrix of some scalar field derivatives \cite{key-8}. If we
want to restrict the Lagrangian to at most two derivatives in each
addend, we construct the scalar density with non-metric 2-rank tensor
field:

\begin{equation}
\mathcal{D}\equiv{\left({\det\left[\sum_{b}{{A_{\alpha}^{b}}\otimes{A_{\beta}^{b}}}\right]}\right)^{2k}}\label{eq:16}
\end{equation}

where $A_{\alpha}^{b}$is a vector field in the index $\alpha$and
$b=1,...,N$ with $N\geq4$in order to make sure that the determinant
of the sum of matrices is not identically zero.

We add a kinetic term for these vector fields that obey DEC:

\begin{equation}
{\mathcal{L}_{IK}^{a}}=-\sum_{b,c}\mu^{bc}{F_{\alpha\beta}^{b}}{g^{c\mu\nu}}{g^{c\beta\nu}}{F_{\mu\nu}^{b}}\sqrt{{g^{a}}}\label{eq:17}
\end{equation}

where ${F_{\alpha\beta}^{b}}\equiv{\partial_{\alpha}}{A_{\beta}^{b}}-{\partial_{\beta}}{A_{\alpha}^{b}}$
and $\mu^{bc}$ are non-negative constants.

The energy momentum tensor that is derived from the kinetic term constitutes
another source in the field equation for the metric $g_{\mu\nu}^{a}$,
and it obeys DEC with respect to this specific metric for all field
configurations. This can be proven for any anti-symmetric field ${F_{\alpha\beta}}$
and interaction in the form of (\ref{eq:17}), as in the case of electromagnetic
field coupled to our standard metric.

The energy momentum tensors (\ref{eq:11}) have to obey asymptotic
conditions $T_{\mu\nu}^{a}\simeq O({r^{-3-\varepsilon}})$in order
to maintain the boundary conditions (\ref{eq:6}) for the metric fields.
For any positive integer $k$ one can assume the desired boundary
conditions on the fields $A_{\alpha}$and its first derivatives so
they go to zero fast enough in order to maintain the required asymptotic
conditions for the energy momentum tensors. These boundary conditions
are consistent with the $A_{\alpha}$ field equtions.

A genenaral multi-graviton interaction for a stable system is then

\begin{equation}
{\mathcal{L}_{I}}=\sum\limits _{k}{\mathcal{L}_{I}^{k}}+\sum_{a}{\mathcal{L}_{IK}^{a}}\label{eq:18}
\end{equation}

where the addends are defined in eq. (\ref{eq:15a}),(\ref{eq:16}),(\ref{eq:17}).
According to the positive energy theorem for interacting multi-metric
systems the Lagrangian in eq.(\ref{eq:1}) with the interaction (\ref{eq:18})
describe a stable multi-graviton theory with non-negative energy,
where the lowest energy state is obtained when all the metric fields
are Minkowskian $g_{\mu\nu}^{a}={\eta_{\mu\nu}}$ and all energy momentum
tensors are zero.

\section{Discussion}

The model presented in this paper, defined by the first and last equation,
is a stable general interacting multi-graviton theory, with at most
two derivative terms is the Lagrangian.

In the construction of this non-negative energy model we did not use
special interactions that remove ghost degrees of freedom, Thus our
model avoids super-luminal shock waves and possible causality problems
which have been claimed for the ghost free single metric massive gravity
and its bi-metric extension, based on that interaction.\cite{key-3}

Furthermore, our model enjoys a strong immunity to causal problems
in general, and not just from those subjected to the above special
constraints. All energy momentum tensors obey DEC with respect to
the metric for which they are the source in the field equations for
that metric. According to Hawking \cite{key-14} and Tipler \cite{key-13},
in general relativity, if the energy momentum tensor obeys the weak
energy condition (WEC) then closed time-like curves (CTC) can be created
in a compact region only if singularities are created also. DEC implies
WEC, therefore CTC in any spacetime described by some metric field,
where time-likeness is with respect to the same metric, can be created
only if there are singularities in that spacetime. Indeed, singularity
in these scenarios does not mean neccesarily a point where the curvature
is infinite. and the connection of the interaction to such a singularity
is less obvious than to infinite curvature. These singularities can
not be excluded simply by regularitiy characters of energy momentum
tensors, as they do have in our model, and in fact there are models
in general relativity with CTC in an asymptotically flat spacetime
region with standard matter and finite curvature \cite{key-15}. However,
these singularity conditions definitely reduce the chance for CTC
in our model, and even if one may construct a stable multi-metric
theory with CTC, it does not constitute evidence for its illness any
more than that the time-machine models mentioned above indicates illness
in general relativity.

Our Lagrangian is composed from a general combination of interactions
that include every subset of the graviton fields, thus the number
of different metric fields on each vertex can take any value. 

The model presented in this paper include the case of N=1, i.e. general
relativity with modified self interaction, but not massive gravity.
Massive terms are quadratic in the graviton fields. In order to maintain
boundary conditions that are needed to define finite total energy
the energy momentum tensors can not include such terms, assuming that
standard matter may also included. Massive terms are either repulsive,
i.e. contribute negative energy density therefore violate DEC, or
with an opposite sign but do not suit appropriate boundary conditions.
Traditionally, multi-metric theories were constructed as a generalisation
of a ghost-free massive gravity. It is shown here explicitly that
the existence of stable massive gravity is not necessary for consistent
interacting multi-gravity theories.

\section{acknowledgments}

I want to thank Lawrence P. Horwitz for careful reading and useful
discussions, to Amos Ori for a detailed and clarifying consultation
about closed time-like curves and singularity theorems, and to Mordehai
Milgrom for his point of view on symmetry breaking vs. consistency.

I would also like to thank Shmuel Elitzur for thorough discussions,
in which many details were modified and explained. 

I gratefully acknowledge financial support from Ariel University. 

This research is supported by the I-CORE Program of the Planning and
Budgeting Committee and the Israel Science Foundation (Grant No. 1937/12)

\end{document}